
\input harvmac

\def\t{\tau}
\def\s{\sigma}
\def\d{\partial}
\def\pder#1{{\partial\over\partial{#1}}}
\def\Pder#1{{\delta\over\delta{#1}}}
\def\PDer#1{{D\over D{#1}}}
\def\T{{1\over4\pi\alpha'}}
\def\l{l_{s}}
\def\intp{\int {d^p\sigma\over (2\pi)^p}}
\def\sump{\sum_{n\in {\bf Z}^p}}

\def\eder{\exp(i\pi B_{ij}
 {{\overleftarrow{\partial}}\over\partial\sigma_i}
 {{\overrightarrow{\partial}}\over\partial\sigma_j})}
\def\edert{\exp(i\pi B_{ab}l_s^2
 {{\overleftarrow{\partial}}\over\partial\tilde{\sigma}_a}
 {{\overrightarrow{\partial}}\over\partial\tilde{\sigma}_b})}
\def\EB{e^{-i\pi B_{ij}m^in^j}}
\def\eb{e^{-i\pi B_{ij}m^in^j}}
\def\Tp{{\bf T}^p}
\def\Tb{{\bf T}^p_B}
\def\tTp{\tilde{\bf T}^p}
\def\ts{\tilde{\sigma}}

\def\Tr{{\rm Tr}}
\def\trn{{\rm tr}_N}
\def\trz{{\rm tr}_{{\bf Z}^p}}
\def\th{\theta}
\def\al{\alpha}
\def\bt{\beta}
\def\ga{\gamma}

\def\np#1#2#3{{ Nucl. Phys.} {\bf B#1}, #2 (19#3)}

\def\pln#1#2#3{Phys. Lett. {\bf B#1}, #2 (19#3)}

\def\pr#1#2#3{{ Phys. Rev.} {\bf D#1}, #2 (19#3)}
\def\prl#1#2#3{{ Phys. Rev. Lett.} {\bf #1}, #2 (19#3) }
\def\ann#1#2#3{{ Ann. Phys.} {\bf #1}, #2 (19#3)}

\def\jmp#1#2#3{{ J. Math. Phys.} {\bf #1}, #2 (19#3)}
\def\hpt#1{{\tt hep-th/#1}}

\lref\Pcski{
J. Polchinski,
``Dirichlet Branes and Ramond-Ramond Charges'',
\prl{75}{4724-4727}{95}, \hpt{9510017}.
}
\lref\Dbrane{
J. Polchinski, S. Chaudhuri, and C. V. Johnson, 
``Notes on D-Branes'',
\hpt{9602052}\semi
J. Polchinski, 
``TASI Lectures on D-Branes'',
\hpt{9611050}.
}
\lref\W{
E. Witten,
``Bound States Of Strings And $p$-Branes'',
\np{460}{335-350}{96}, \hpt{9510135}.
}
\lref\GQM{
M. Baake, P. Reinicke, and V. Rittenberg,
``Fierz identities for real Clifford algebras and the number 
of supercharges'',
\jmp{26}{1070-1071}{85}\semi
R. Flume,
``On Quantum Mechanics with Extended Supersymmetry 
and Nonabelian Gauge Constraints'',
\ann{164}{189-220}{85}\semi
M. Claudson and M. B. Halpern,
``Supersymmetric Ground State Wave Functions'',
\np{250}{689-715}{85}.
}
\lref\BFSS{
T. Banks, W. Fischler, S. H. Shenker, and L. Susskind,
``M Theory As A Matrix Model: A Conjecture'',
\pr{55}{5112-5128}{97}, \hpt{9610043}.
}
\lref\IKKT{
N. Ishibashi, H. Kawai, Y. Kitazawa, and A. Tsuchiya,
``A Large N Reduced Model as Superstring'',
\np{498}{467-491}{97}, \hpt{9612115}.
}
\lref\DH{
M. R. Douglas and C. Hull,
``D-Branes and the Noncommutative Torus'',
\hpt{9711165}.
}
\lref\CDS{
A. Connes, M. R. Douglas, and A. Schwarz,
``Noncommutative Geometry and Matrix Theory: 
Compactification on Tori'',
\hpt{9711162}.
}
\lref\Tay{
W. Taylor,
``D-brane field theory on compact spaces'',
\pln{394}{283-287}{97}, \hpt{9611042}.
}
\lref\HWW{
P.-M. Ho, Y.-Y. Wu, and Y.-S. Wu,
``Towards a Noncommutative Geometric Approach to Matrix
Compactification'',
\hpt{9712201}.
}
\lref\HW{
P.-M. Ho and Y.-S. Wu,
``Noncommutative Gauge Theories in Matrix Theory'',
\hpt{9801147}.
}
\lref\Li{
M. Li,
``Comments on Supersymmetric Yang-Mills Theory on a Noncommutative
Torus'',
\hpt{9802052}
}
\lref\Berk{
M. Berkooz,
``Non-local Field Theories and the Non-commutative Torus'',
\hpt{9802069}.
}
\lref\Casa{
R. Casalbuoni,
``Algebraic treatment of compactification on noncommutative Tori'',
\hpt{9801170}.
}
\lref\LR{
R. G. Leigh and M. Rozali,
``A Note on Six-Dimensional Gauge Theories'',
\hpt{9712168}
}
\lref\Shen{
S. H. Shenker,
``Another Length Scale in String Theory?'',
\hpt{9509132}
}
\lref\DKPS{
M. R. Douglas, D. Kabat, P. Pouliot, and S. H. Shenker,
``D-branes and Short Distances in String Theory'',
\np{485}{85-127}{97}. 
}
\lref\GRT{
O. J. Ganor, S. Rangoolam, and W. Taylor,
``Branes, Fluxes and Duality in M(atrix)-Theory'',
\np{492}{191-204}{97}, \hpt{9611202}. 
}
\lref\Roz{
M. Rozali,
``Matrix Theory and U-Duality in Seven Dimensions'',
\pln{400}{260-264}{97}, \hpt{9702136}.
}
\lref\BRS{
M. Berkooz, M. Rozali, and N. Seiberg,
``Matrix Description of M theory on $T^4$ and $T^5$'',
\pln{408}{105-110}{97}, \hpt{9704089}.
}
\lref\new{
N. Seiberg,
``New Theories in Six Dimensions and Matrix Description of M-theory 
on $T^5$ and $T^5/Z_2$'',
\pln{408}{98-104}{97}, \hpt{9705221}.
}
\lref\Sen{
A. Sen, 
``D0 Branes on $T^n$ and Matrix Theory'',
\hpt{9709220}.
}
\lref\why{
N. Seiberg,
``Why is the Matrix Model Correct?'',
\prl{79}{3577-3580}{97}, \hpt{9710009}.
}
\lref\Dzero{
Y.-K. E. Cheung and M. Krogh,
``Noncommutative Geometry from 0-branes in a Background B-field'',
\hpt{9803031}
}

\Title{                                \vbox{\hbox{UT-810}
                                             \hbox{\tt hep-th/9803044}} }
{\vbox{\centerline{
              Matrix Theory on Noncommutative Torus 
}}}

\vskip .2in

\centerline{
               Teruhiko Kawano and Kazumi Okuyama
}

\vskip .2in 

\centerline{
               Department of Physics, University of Tokyo
}
\centerline{
               Hongou, Tokyo 113-0033, Japan
}
\centerline{\tt
              kawano@hep-th.phys.s.u-tokyo.ac.jp
}
\vskip -1mm
\centerline{\tt
              okuyama@hep-th.phys.s.u-tokyo.ac.jp
}

\vskip 3cm

We consider the compactification of Matrix theory on tori with 
background  antisymmetric tensor field. 
Douglas and Hull have recently discussed 
how noncommutative geometry appears on the tori. 
In this paper, we demonstrate the concrete construction of 
this compactification of Matrix theory 
in a similar way to that previously given by Taylor.

\Date{March, 1998}

\newsec{Introduction}

In the recent development of string theory, D-branes played the essential 
role\Pcski\ (See also \Dbrane\ for references therein). 
One of the striking features of D-branes is that the 
transverse coordinates of D-branes are
promoted to matrices \W.
Therefore, spacetime for D-branes is noncommutative. 
However, in the situations which have been discussed so far, 
the coordinates parallel to the D-brane
worldvolume were 
commutative quantities. We are thus led to ask 
whether or not the D-brane worldvolume can also become noncommutative.

In \BFSS,
it was conjectured that M-theory is microscopically described 
by Matrix theory;
$(0+1)$-dimensional matrix quantum mechanics \GQM\ which is obtained by the
dimensional reduction of $D=10$ $N=1$ super Yang-Mills theory.
The action of Matrix theory 
can be also interpreted as the low-energy effective action of D0-branes.

Another Matrix theory, known as the IKKT model, was proposed \IKKT\ to give 
a nonperturbative formulation of type IIB superstring theory. 
Although our formulation in this paper can be similarly applied to 
the IKKT model, in order to make our discussion simple, we will 
restrict ourselves to the former theory proposed in \BFSS. 

Matrix theory compactified on a torus 
is described by super Yang-Mills theory on the dual 
torus \refs{\BFSS,\Tay,\GRT} if the dimension of the torus is less than four 
(see \refs{\Roz\BRS\new\Sen{--}\why} for the compactification on higher dimensional
tori). 
Especially, in \Tay, Taylor has presented the clear picture
of the compactification of Matrix theory on a torus.
He showed that by putting the D0-branes and their images under the
lattice translations on the covering space,
super Yang-Mills theory is obtained from the original
D0-brane action by the Fourier transformation.  

More recently, the compactification of Matrix theory 
on a noncommutative torus was
considered \refs{\CDS\HWW\HW{--}\Casa}. In \CDS, it was discussed that
this theory 
corresponds 
to the compactification of M-theory on a
torus with constant background three-form  field and  is
described  by the super Yang-Mills theory on the dual
noncommutative torus. See \refs{\Li,\Berk,\LR} for the related issues.

Douglas and Hull have also shown\DH\ that super Yang-Mills theory
on the noncommutative torus  naturally appears as the
D-brane worldvolume theory, and the coordinates of the 
D-brane worldvolume can
become noncommutative. 
Instead of considering D0-branes on a two-dimensional torus 
with constant background NS two-form field, they considered D1-branes on the slanted torus obtained by performing T-duality along one direction of the torus 
and showed that the typical interaction of gauge theory on 
the noncommutative torus can appear in this system.

As is mentioned in \DH, it is
possible to understand the appearance of noncommutativity directly in the
original D0-brane picture.  In this paper, 
starting from the action of D0-branes, we derive 
super Yang-Mills theory on the dual noncommutative torus 
by extending the method of \Tay\ to the compactification with
background NS two-form field.

This paper is organized as follows: 
In section 2, 
we will study the wave function of an open string with the Dirichlet boundary
condition
in the constant background NS two-form field 
and show how it is transformed when we locally gauge away the
NS two-form field. 
In section 3, we will study the action of D0-branes on the torus
with the background NS two-form field, 
and, following \Tay, derive 
super Yang-Mills theory on the dual noncommutative torus. 
Section 4 is devoted to discussion.

\newsec{D0-Branes on Torus with Background Antisymmetric Tensor}

D-branes are described by using open strings with the Dirichlet boundary 
conditions\Dbrane. We will consider D0-branes on a torus $\Tp$.
The coordinates of the torus $\Tp$ have the following identification:
\eqn\torus{
X^i \sim X^i + 2\pi\l m^i, \quad (i=1,\cdots, p)
}
with $m^i\in{\bf Z}$ and $\alpha'=(\l)^2$. 
The string tension is $T={1\over2\pi\alpha'}$.
In this section, since the string coordinates which describe the torus $\Tp$ 
are relevant to our discussion, we focus our attention on the following part 
of the action: 
\eqn\S{\eqalign{
S&=\T\int d\t \int^{\pi}_{0}d\s \left[G_{ij}\d_{\alpha}X^i\d^{\alpha}X^j
+\epsilon^{\alpha\beta}B_{ij}\d_{\alpha}X^i\d_{\beta}X^j\right]
\cr
&=\int d\t L\left[X, \d_0X\right],
\cr
}}
where $G_{ij}$ and $B_{ij}$ are constant background metric and 
NS two-form field, respectively. 

The conjugate momenta $P_i(\s)$ of the string coordinates $X^i(\s)$ 
are defined in the usual way;
\eqn\P{\eqalign{
P_i(\s)&={\delta\over\delta \d_0X(\s)} L
\cr
       &={1\over2\pi\alpha'}\left[G_{ij}\d_0X^i(\s)+B_{ij}
\int_0^{\pi}d\s'\delta(\s,\s')\d_1X^j(\s')\right]
\cr
       &=-i\Pder{X^i(\s)}.
\cr
}}
where $\delta(\s,\s')=\sum_{k=1}^{\infty}2/\pi\sin(k\s)\sin(k\s')$ is
the $\delta$-function 
on the space of functions with the Dirichlet boundary condition.
Note that the integral including $\delta(\s,\s')$ in eq.\P\
is not equal to $\d_1X^j(\s)$,
 because $\d_1X^j$ does not satisfy the Dirichlet
boundary condition.
Then we obtain the relevant part of the Hamiltonian
\eqn\H{\eqalign{
&H=\int^{\pi}_{0}d\s \d_0X^iP_{i}(\s) - L
\cr
&={1\over2}\int^{\pi}_{0}d\s\left[
-(2\pi\alpha')\PDer{X^i(\s)}G^{ij}\PDer{X^j(\s)}
+{1\over2\pi\alpha'}\d_1X^i(\s)G_{ij}\d_1X^j(\s)\right],
\cr
}}
where
\eqn\PD{
\PDer{X^i(\s)}=\Pder{X^i(\s)}-{i\over2\pi\alpha'}B_{ij}\int_0^{\pi}d\s'
\delta(\s,\s')\d_1X^j(\s').
}
The wavefunction $\Psi[X]$ of this system should obey the Schr\"odinder 
equation; $i\pder{\t}\Psi=H\Psi$.
If we introduce the following operator:
\eqn\U{
U=\exp\left[{i\over4\pi\alpha'}\int_0^{\pi} d\s B_{ij}X^i(\s)\pder{\s}X^j(\s)\right]
}
which satisfies that
\eqn\unitary{
\PDer{X^i(\s)}=U \Pder{X^i(\s)} U^{-1}
}
and if we define another wavefunction $\Psi^{(0)}$ by
\eqn\Utrf{
\Psi=U\Psi^{(0)},
}
this wavefunction $\Psi^{(0)}$ has the familiar Hamiltonian $H_0$ 
with no background tensor $B_{ij}$; namely, the Hamiltonian $H$ 
where $\PDer{X^i(\s)}$ is replaced by $\Pder{X^i(\s)}$.

Now we would like to consider an open string with
the Dirichlet boundary condition
\eqn\Dbc{\bigg\{\matrix{
X^i(\s=0)&=2\pi\l n^i,
\cr
X^i(\s=\pi)&=2\pi\l m^i.
\cr}
}
We will call this sector the $(m,n)$ sector.

In the $(m,n)$ sector, the wavefunction $\Psi_{m,n}[X]$ is related by 
the unitary transformation \Utrf\ to the wavefunction $\Psi^{(0)}_{m,n}[X]$ 
which has the Hamiltonian $H_0$ with no background two-form field; 
$\Psi_{m,n}[X]=U\Psi^{(0)}_{m,n}[X]$, 
as we discussed above.
Similarly, in the $(m-n,0)$ sector, $\Psi_{m-n,0}[X]=U\Psi^{(0)}_{m-n,0}[X]$.
If we assume that $\Psi^{(0)}_{m,n}[X]=\Psi^{(0)}_{m-n,0}[X]$, 
then we obtain that 
\eqn\key{
\Psi_{m,n}[X]=\EB\Psi_{m-n,0}[X].
}
Thus in the relation of the wavefunction in the $(m,n)$ sector with 
that in the $(m-n,0)$ sector, we have the phase factor $\EB$, 
which will account for the noncommutativity of this torus 
in Matrix theory.

For later convenience, we introduce the vielbein 
\eqn\bein{
G_{ij}={E^a}_i{E^b}_j\delta_{ab}.
}
We change the indices by multiplying the vielbein ${E^a}_i$;
\eqn\atoi{\eqalign{
&X^a={E^a}_iX^i, \cr
&B_{ab}=B_{ij}{E^i}_a{E^j}_b,
\cr}
}
where ${E^i}_a$ is the inverse of the vielbein ${E^a}_i$.
The periodicity \torus\ of $X^i$ is rewritten in $X^a$ as
\eqn\Xaperiod{
X^a \sim X^a+2\pi\l{E^a}_im^i.
}

\newsec{Matrix Theory Compactified on Noncommutative Torus}

In this section, we consider the compactification of Matrix theory on the 
torus with the background NS two-form field.  
The action of Matrix theory\BFSS\ is given in the string metric by
\eqn\matact{\eqalign{
S=T_0\int dt~\Tr\Biggl\{&
{1\over 2}(D_tX^I)^2+{T^2\over 4}[X^I,X^J]^2 \cr
&+{i\over 2}\th^{\al}D_t\th^{\al}+{T\over 2}\th^{\al}\ga^I_{\al\bt}
[X_I,\th^{\bt}]
\Biggr\}. 
\cr}
}
where $T_0=1/(g_s\l)$ is the mass of D0-branes.

To obtain the compactification of Matrix theory on 
$\Tp \simeq {\bf R}^p/{\bf Z}^p$, 
we put D0-branes and their images under the lattice translation 
${\bf Z}^p$ on the covering space ${\bf R}^p$  \Tay.
If we put $N$ D0-branes per unit cell of the lattice, then the Chan-Paton 
factor is labeled by $(m,i)$ where $m\in {\bf Z}^p$ and $i=1,\ldots,N$.
The dynamical variables of this theory
are nine ordinary matrices $X^I_{(m,i),(n,j)}~(I=1,\ldots,9)$ and
sixteen Grassmann matrices $\th^{\al}_{(m,i),(n,j)}~(\al=1,\ldots,16)$.
Henceforth, we will suppress the $N$-indices $i,j$, for simplicity. 

Furthermore, following Taylor \Tay, 
to specify the compactification on $\Tp$, 
we should impose on these variables $X^I$, $\th^{\al}$ the following relation: 
\eqn\Bzeroq{\eqalign{
&X^{A}_{m,n}=X^{A}_{m-n,0} \quad (A=p+1,\cdots,9),  
\cr
&X^{a}_{m,n}=X^{a}_{m-n,0}
 +2\pi\l E^a_jn^j\delta_{m,n} \quad (a=1,\cdots,p), 
\cr
&\th^{\al}_{m,n}=\th^{\al}_{m-n,0}.
\cr}
}

Now we turn on the background $B_{ij}$ 
on the torus. The variables $X^I$, $\th^{\al}$ can be interpreted in 
terms of the D0-brane low-energy effective action as the Higgs field and 
their superpartner, respectively, which come from the massless modes of 
open string which has its ends on D0-branes \W. 
Therefore, as we have learned in eq.\key\ of the previous section, 
it seems natural that, to specify our torus with the background $B_{ij}$, 
we should replace the defining relation \Bzeroq\ with 
\eqn\reducing{\eqalign{
&X^{A}_{m,n}=\eb X^{A}_{m-n,0} ~~~(A=p+1,\cdots,9),  
\cr
&X^{a}_{m,n}=\eb (X^{a}_{m-n,0}
 +2\pi\l E^a_jn^j\delta_{m,n}) ~~~(a=1,\cdots,p), 
\cr
&\th^{\al}_{m,n}=\eb \th^{\al}_{m-n,0}.
\cr}}

To go to the T-dual picture, as it was done in \Tay,
we make the Fourier transformation 
from $n\in {\bf Z}^p$ to $\s\in [0,2\pi]^p$; 
$X(\sigma)=\sump e^{in^j\sigma_j}X_{n,0}$, 
$\th(\s)=\sump e^{in^j\sigma_j}\th_{n,0}$.
These variables $X(\s)$ and $\th(\s)$ are 
$N\times N$ hermitian matrix-valued functions.
The commutators of the matrix variables are expressed in terms of 
these variables $X(\s)$, $\th(\s)$ as 
\eqn\comXX{\eqalign{
[X^A,X^B]_{m,n}&
=\eb\intp e^{-i(m-n)\s}\{X^A,X^B\}_B(\s) \cr
&=\eb \bigl(\{X^A,X^B\}_B\bigr)_{m-n},\cr
[X^A,\th^{\al}]_{m,n}&=\eb \bigl(\{X^A,\th^{\al}\}_B\bigr)_{m-n},
\cr}}
and
\eqn\comXab{\eqalign{
[X^a,X^A]_{m,n}&=\eb {1\over i}
  \bigl(2\pi\l E^a_i\d^iX^A+i\{X^a,X^A\}_B\bigr)_{m-n},
\cr
[X^a,\th^{\al}]_{m,n}&=\eb {1\over i}
  \bigl(2\pi\l E^a_i\d^i\th^{\al}+i\{X^a,\th^{\al}\}_B\bigr)_{m-n},\cr
[X^a,X^b]_{m,n}&=\eb {1\over i}
 \bigl(2\pi\l E^a_i\d^iX^b-2\pi\l E^b_i\d^iX^a
 +i\{X^a,X^b\}_B\bigr)_{m-n},
\cr}
}
where $\{~,~\}_B$ stands for the Moyal bracket.
The Moyal bracket is the commutator of two functions by $*$-product
\eqn\moyaldef{
\{f,g\}_B \equiv (f*g)_B-(g*f)_B,
}
where the $*$-product is defined by
\eqn\star{
(f*g)_B(\s) \equiv f(\s)\eder g(\s), 
}
as for the $N$-indices, which should be understood 
as $(f*g)_{ij}=f_{ik}*g_{kj}$.

We rescale the variable $X^a(\s)$ into $A^a(\s)$;   
\eqn\Ala{
A^a(\s)=TX^a(\s)={1\over 2\pi\l^2}X^a(\s),
}
and the parameters $\s_i$ into the ``dual coordinates'' $\ts_a$;
$\ts_a=\l E_a^i\s_i.$ Note that the dual coordinates $\ts_a$ have 
the periodicity; $\ts_a \sim \ts_a+2\pi\l E_a^i$ 
and thus have the dual relation to the identification \Xaperiod\ for $X^a$. 
The $*$-product is rewritten in the dual coordinates $\ts_a$ as
\eqn\starts{
(f*g)_B(\ts)=f(\ts)\edert g(\ts).
}

Defining the covariant derivative and the field strength by
\eqn\covD{\eqalign{
&D^af(\ts)=\pder{\ts_a}f(\ts)+i\{A^a,f\}_B(\ts),
\cr
&F^{ab}(\ts)=\pder{\ts_a}A^b(\ts)-\pder{\ts_b}A^a(\ts)+i\{A^a,A^b\}_B(\ts),
\cr}
} 
in the end, for the compactification on the torus with the background $B_{ij}$, 
we find the action 
\eqn\SYM{\eqalign{
S=M_p\int dtd^p\tilde{\s}\ \trn\Biggl\{&
-{1\over 4T^2}F_{\mu\nu}F^{\mu\nu} 
-{1\over2}(D_{\mu}X^A)^2+{T^2\over4}\bigl(\{X^A,X^B\}_B\bigr)^2
\cr
&\quad-{i\over 2}\th^{\al}\ga^{\mu}_{\al\bt}D_{\mu}\th^{\bt}
+{T\over 2}\th^{\al}\ga^A_{\al\bt}
 \{X_A,\th^{\bt}\}_B
\Biggr\},
\cr}
}
with $\mu,\nu=t,1,\ldots,p$ and $\ga^0_{\al\bt}=-\delta_{\al\bt}$, 
where the integration is over ${{\bf R}\times \tTp}$, and we also 
defined $M_p=T_p\sqrt{\det G_{ij}}$ with $T_p=T_0(2\pi\l)^{-p}$ 
the tension of Dp-branes \Pcski. Note that we have divided the action by 
the irrelevant factor $\trz({\bf 1})=\sump 1$ which came from the lattice 
translation.

\newsec{Discussion}

In this paper, we have considered the compactification of Matrix theory 
on the torus with the background antisymmetric tensor. 
We have followed the method for the compactification on the torus given 
by Taylor\Tay\ and extended it to apply to our torus.
We have thus found that the resultant action \SYM\ is that formally obtained 
from the standard action of super Yang-Mills theory by replacing 
the ordinary product of functions by the $*$-product. 
This fact shows that \SYM\ can be thought of
as the action of super Yang-Mills theory on the noncommutative torus 
$\Tb$ which is associated with the algebra defined by
\eqn\alg{
Z_iZ_j=e^{-2\pi iB_{ij}}Z_jZ_i
}
with the identification $Z_i \sim e^{i\s_i}$, 
as Connes, Douglas, and Schwarz have already discussed \CDS.

Note that, as pointed out in \DH, in connection with 
the issue of the minimal length, the radii of the dual torus $\tTp$ on which
super Yang-Mills theory is defined are the radii obtained by performing 
T-duality for the torus without the background NS two-form.
This phenomenon is related to the fact that, in the NS two-form background, 
open strings behave differently from closed strings. 
It is important to clarify such differences in more general background
for our understanding of the compactification of Matrix theory.

The authors of \CDS\ have discussed  the relation between super Yang-Mills 
theory on the noncommutative torus and M theory on tori with constant 
background three-form field.
In \DH, Douglas and Hull have also discussed the appearance 
of the noncommutative geometry on tori with constant background 
two-form field in Type II superstring theory.
Therefore, we believe that the significance of our paper is 
to give the simple derivation and to demonstrate 
the compactification of Matrix theory on such tori 
from the point of the simple view given by Taylor \Tay. 

\vskip 0.5cm
\noindent
{\sl Note added}: After this work was completed, we received a paper
\Dzero\ which has overlaps with ours.

\medskip
\centerline{{\bf Acknowledgements}}
K.O. is supported in part by JSPS Research Fellowships for Young
Scientists.

\listrefs

\end